\documentclass[graybox,natbib,nosecnum]{svmult}
\bibpunct{(}{)}{;}{a}{}{,} 

\pdfoutput=1   

\usepackage{mathptmx}       
\usepackage{helvet}         
\usepackage{courier}        
\usepackage{type1cm}        

\usepackage{makeidx}         
\usepackage{graphicx}        
\usepackage{multicol}        
\usepackage[bottom]{footmisc}
\usepackage[normalem]{ulem}	
\usepackage{hyperref}  
\usepackage{url}  

\usepackage{soul}   

\makeindex             

\begin{document}

%
\def\ltsima{$\; \buildrel < \over \sim \;$}
\def\lsim{\lower.5ex\hbox{\ltsima}}
\def\gtsima{$\; \buildrel > \over \sim \;$}
\def\gsim{\lower.5ex\hbox{\gtsima}}

\title*{Planet Occurrence: Doppler and Transit Surveys}
\titlerunning{Planet Occurrence Rates}
\author{Joshua N.\ Winn}
\institute{Department of Astrophysical Sciences, Princeton University, 
  4 Ivy Lane, Princeton, NJ 08544, USA, \email{jnwinn@princeton.edu}}
\maketitle

\abstract{Prior to the 1990s, speculations about the occurrence of
  planets around other stars were based only on planet formation
  theory, observations of circumstellar disks, and the knowledge that
  at least one seemingly ordinary star had managed to make a variety
  of different planets.  Since then, Doppler and transit surveys have
  revealed the population of planets around other Sun-like stars,
  especially those with orbital periods shorter than a few years.
  Over the last decade these surveys have risen to new heights with
  Doppler spectrographs capable of 1~m~s$^{-1}$ precision, and space
  telescopes capable of detecting the transits of Earth-sized planets.
  This article is a brief introductory review of the knowledge of
  planet occurrence that has been gained from these surveys.}
\section{Introduction}

If, in some cataclysm, all our knowledge of exoplanets were to be
destroyed, and only one sentence passed on to the next generation of
astronomers, what statement would contain the most helpful
information?\footnote{Adapted from book I, chapter 1, verse 2 of \citet{Feynman1963}.}
Here is one possibility: {\it
  Most Sun-like stars have planets, which display a wider range of
  properties --- size, mass, orbital parameters --- than the planets
  of the Solar System.}

To be quantitative we could give the fraction of stars with planets,
restricted to the types of planets we have managed to detect.  But
since this fraction is so close to unity, it might be more helpful to
specify the average number $n$ of planets per star: the total number
of planets in the galaxy divided by the number of stars.  Better still
would be a mathematical function to show how occurrence depends on the
planet's properties. For example, we could supply a functional form
for
\begin{equation}
\Gamma_{R,P} = \frac{\partial^2n}{\partial\log R~\partial\log P},
\end{equation}
the average number of planets per star per log-intervals in
radius and period.  The number $n$ is the {\it occurrence rate} and
the function $\Gamma_{R,P}$ is an {\it occurrence rate
  density}.

Occurrence rates depend on other planetary parameters, such as orbital
eccentricity, and on the characteristics of the star, such as mass and
metallicity.  Occurrence is also conditional on the properties of any
other planets known to exist around the same star.  No simple function
could account for all these parameters and their correlations.
Ideally, we would transmit a computer program that produces
random realizations of planetary systems that are statistically
consistent with everything we have learned from planet surveys.  This
would help our descendants design new instruments to detect planets,
and inspire their theories for planet formation.

What follows is an introductory review of the progress toward this
goal that has come from Doppler and transit surveys.  The basics of
the Doppler and transit methods themselves are left for other reviews,
such as those by \citet{LovisFischer2010}, \citet{Winn2010}, and
Wright (this volume).  Here we will simply remind ourselves of the
key properties of the Doppler and transit signals:
\begin{eqnarray}
  K & = & \frac{0.64~{\rm m~s}^{-1}}{\sqrt{1-e^2}} \left( \frac{P}{1~{\rm day}} \right)^{-1/3}
  \frac{(M/M_\oplus)\sin I}{(M_\star/M_\odot)^{2/3}}, \\
  \delta & = & 8.4\times 10^{-5}
   \left( \frac{R/R_\star}{R_\oplus/R_\odot} \right)^2,~~p_{\rm tra} = \frac{0.0046}{1-e^2}~\frac{R_\star/a}{R_\odot/1~{\rm AU}} 
\end{eqnarray}
where $K$ is the radial-velocity semiamplitude; $\delta$ is
the fractional loss of light during transits; $p_{\rm tra}$ is the
probability for a randomly-oriented orbit to exhibit transits; $a$,
$P$, $e$, and $I$ are the orbital semimajor axis, period,
eccentricity, and inclination; $M$ and $R$ are the mass and radius of the planet,
and $M_\star$ and $R_\star$ are those of the star.

The next section describes methods for occurrence calculations.
Because the surveys have shown major differences in occurrence between
giant planets and small planets, with a dividing line just above
4~$R_\oplus$ or 20~$M_\oplus$, the results for giants and small
planets are presented separately.  After that comes a review of what
is known about other types of stars, followed by a discussion of
future prospects.

\section{Methods}
\label{sec:methods}

\begin{figure}
\includegraphics[scale=.48]{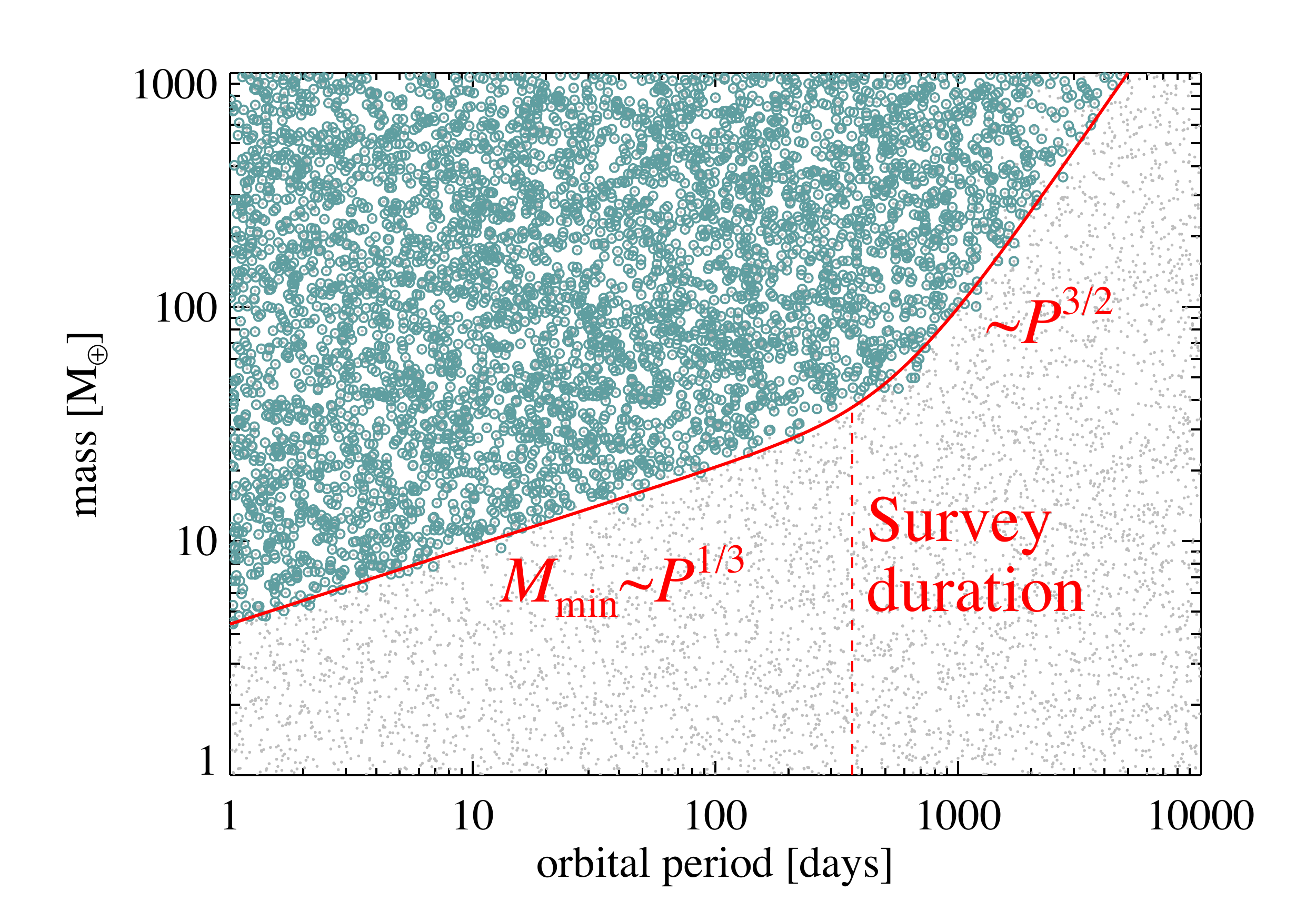}
\caption{{\bf Idealized Doppler survey} of $10^4$ identical Sun-like
  stars. Each star has one planet on a randomly-oriented circular
  orbit, with a mass and period drawn from log-uniform distributions
  between the plotted limits. Each star is observed 50 times over one
  year with 1~m~s$^{-1}$ precision. The small gray dots are all the
  planets; the blue circles enclose those detected with 10$\sigma$
  confidence. For periods shorter than the survey duration, the
  threshold mass is proportional to $P^{1/3}$, corresponding to a
  constant Doppler amplitude.  For longer periods, the threshold mass
  increases more rapidly, with an exponent depending on the desired
  false-alarm probability \citep{Cumming2004}.
  \label{fig:doppler}}
\end{figure}

\begin{figure}
\includegraphics[scale=.48]{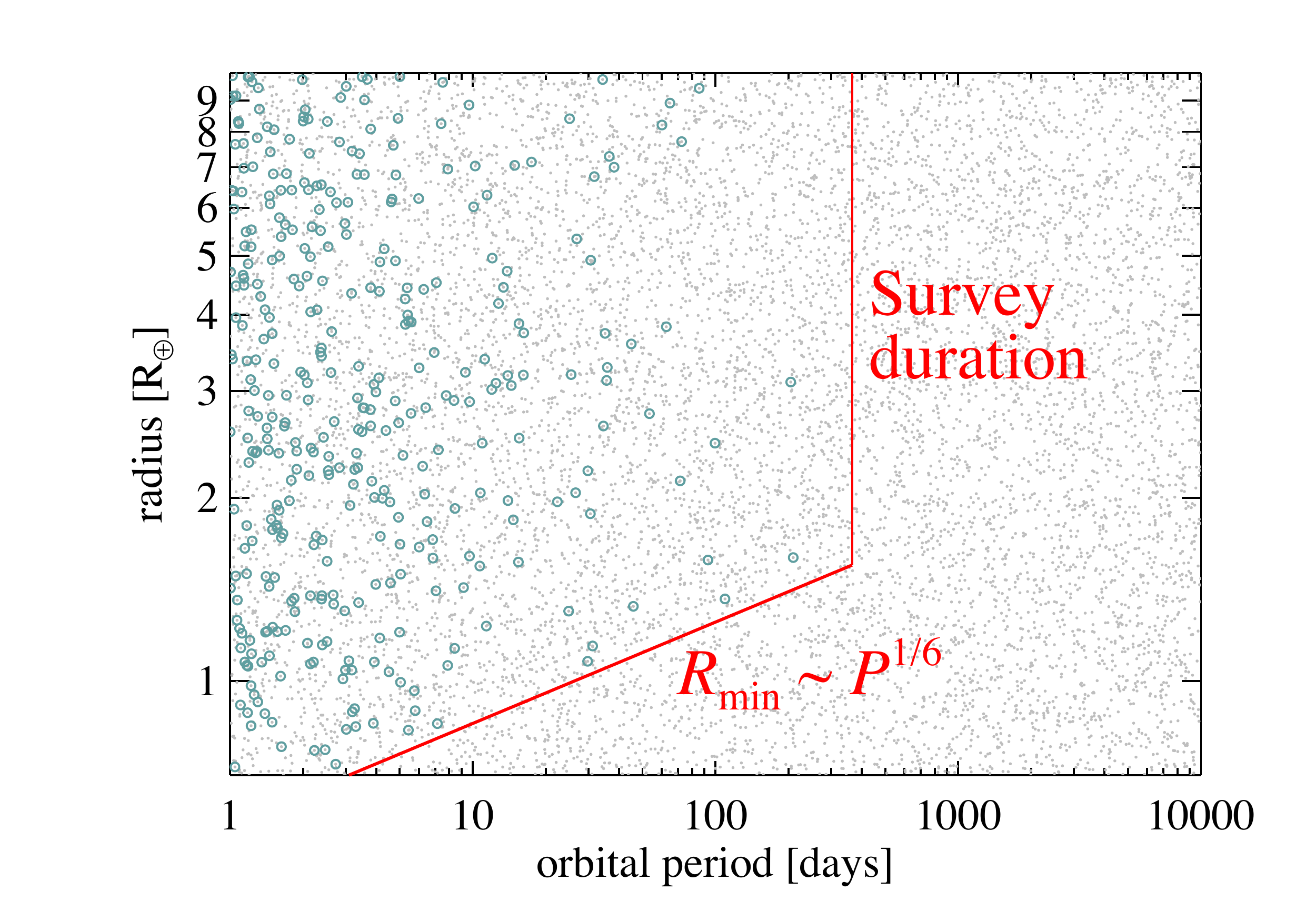}
\caption{{\bf Idealized transit survey} of $10^4$ identical Sun-like
  stars. Each star has one planet on a randomly-oriented circular
  orbit, with a radius and period drawn from log-uniform distributions
  between the plotted limits. Each star is observed continuously for
  one year with a photon-limited photometric precision corresponding
  to $3\times 10^{-5}$ over 6~hours. The small gray dots are all the planets;
  the blue circles enclose those detected with 10$\sigma$ confidence
  based on at least two transit detections. Compared to the Doppler
  survey, the transit survey finds fewer planets and is more strongly
  biased toward short periods, because of the geometric transit
  probability is low and proportional to $P^{-2/3}$.  For orbital
  periods shorter than survey duration, the threshold radius varies as
  $P^{1/6}$ \citep{Pepper+2003}. For longer periods it is impossible
  to observe more than one transit.
  \label{fig:transit}}
\end{figure}

Life would be simple if planets came in only one type and we could
detect them unerringly.  We would search $N$ stars, detect $N_{\rm
  det}$ planets, and conclude $n \approx N_{\rm det}/N$.  But
detection is not assured: small signals can be lost in the noise.  If
the detection probability were $p_{\rm det}$ in all cases, then
effectively we would only have searched $p_{\rm det}N$ stars, and the
estimated occurrence rate would be $N_{\rm det}/(p_{\rm det}N)$.

In reality, $p_{\rm det}$ depends strongly on the characteristics of
the star and planet (see Figures~\ref{fig:doppler} and
\ref{fig:transit}).  Detection is easier for brighter stars, shorter
orbital periods, and larger planets relative to the star.  For this
reason we need to group the planets according to orbital period and
other salient characteristics for detection: the radius $R$, for
transit surveys; and $m \equiv M\sin I$ for Doppler surveys.  Then our
estimate becomes
\begin{equation}
  n_i \approx \frac{N_{{\rm det},i}} {\sum_{j=1}^N p_{{\rm det},ij}},
\end{equation}
where the index $i$ refers to a group of planets sharing the same
characteristics, and the index $j$ specifies the star that was searched.
Transit surveys have the additional problem that most planets
produce no signal at all, because their orbits are not viewed at high
enough inclination.  Thus we must also divide by the probability
$p_{\rm tra}$ for transits to occur.

This conceptually simple method has been the basis of many
investigations. The results of Doppler surveys are presented as a
matrix of occurrence rates for rectangular regions in the space of
$\log m$ and $\log P$; for transit surveys the regions are in the
space of $\log R$ and $\log P$.  Ideally, each region is large enough
to contain many detected planets, and yet small enough that the
detection probability does not vary too much from one side to the
other.

In practice these conditions are rarely achieved, and other methods
are preferred. One approach is to posit a parameterized
functional form for the occurrence rate density, such as a power law
\begin{equation}
\label{eqn:powerlaw}
\Gamma_{m,P} = \frac{\partial^2n}{\partial\log m~\partial\log P} = C\,m^\alpha P^\beta,
\end{equation}
and use it to construct a likelihood function for the outcome of a
survey.  This function must take into account the detection
probability, the properties of the detected systems, and the
properties of the stars for which no planets were detected.  Then the
values of the adjustable parameters are determined by maximizing the
likelihood.  Details are provided by \citet{TabachnikTremaine2002},
\citet{Cumming+2008}, and \citet{Youdin2011}.
\citet{ForemanMackey+2014} cast the problem in the form of Bayesian
hierarchical inference, emphasizing the importance of accounting for
observational uncertainties in the planet and stellar properties.

Most studies report the occurrence rate density as a function of
planet properties, regardless of any other planets in the system. It
is more difficult to quantify the {\it multiplicity} of planetary
systems, the number of planets that orbit together around the same
star.  For Doppler surveys, one problem is that the star is pulled by
all the planets simultaneously.  As a result, the detectability of a
given planet depends on the properties of any other planets ---
especially their periods --- and on the timespan and spacing between
the data points.  This makes it difficult to calculate the detection
probabilities.  For transit surveys, the overlap between different
planetary signals is minimal; instead the problem is a degeneracy
between multiplicity and inclination dispersion.  A star with only one
detected planet could lack additional planets, or it could have several
planets only one of which happens to transit.  In principle this
degeneracy can be broken by combining the results of Doppler and
transit surveys \citep{TremaineDong2012}.

Doppler surveys have uncovered a total of about 500 planets. The most
informative surveys for planet occurrence were based on observations
with the High Resolution Echelle Spectrometer (HIRES) on the Keck~I
10-meter telescope \citep{Cumming+2008,Howard+2010} and the High
Accuracy Radial-velocity Planet Searcher (HARPS) on the La Silla
3.6-meter telescope \citep{Mayor+2011}. Both instruments were used to
monitor $\sim$$10^3$ stars for about a decade, with a precision of a
few meters per second.  Additional information comes from a few
lower-precision and longer-duration surveys \citep[see,
  e.g.,][]{LovisFischer2010}.

For transits, the ground-based surveys have discovered about 200
planets, but are not well suited to occurrence calculations because
the sample of searched stars and the detection probabilities are
poorly characterized.  Instead our most important source is the NASA
{\it Kepler} mission, which used a 1-meter space telescope to measure
the brightness of 150,000 stars every 30 minutes for 4 years
\citep{Borucki2016}.  The typical photometric precision over a 6-hour
time interval was of order $10^{-4}$.  This was sufficient to detect
several thousand planets.

\section{Giant planets}
\label{sec:giant}

\begin{figure}
\includegraphics[scale=.50,angle=0]{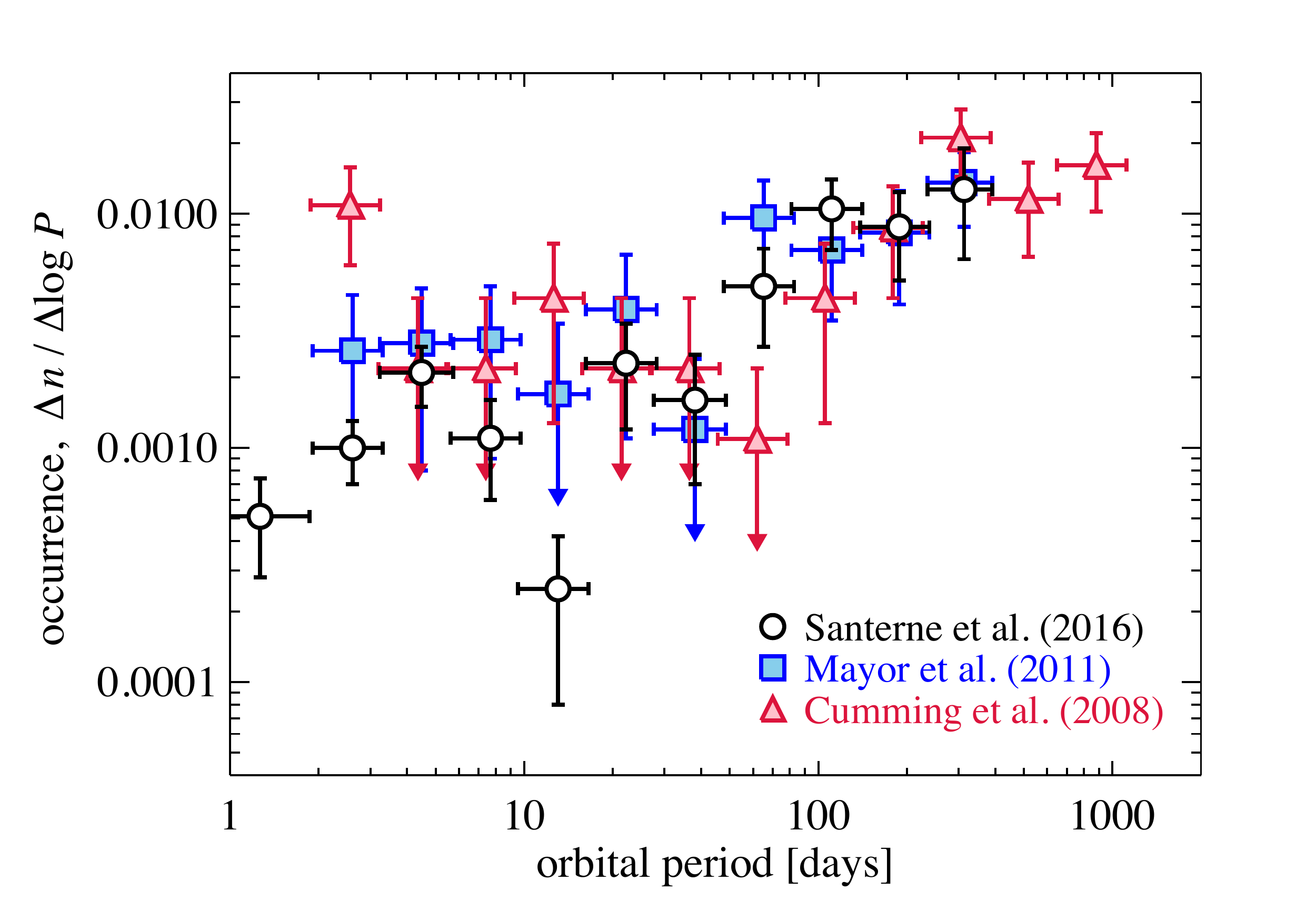}
\caption{Occurrence of giant planets as a function of orbital period,
  from the {\it Kepler} transit survey and two independent Doppler
  surveys.  Data from \citet{Cumming+2008} refer to planets with
  minimum mass $m>0.3\,M_{\rm Jup}$; data from \citet{Mayor+2011} are
  for planets with $m>0.16\,M_{\rm Jup}$; data from
  \citep{Santerne+2016} are for planets with a radius in the
  approximate range 0.5--2\,$R_{\rm Jup}$. Rates are reported in
  planets per star per $\Delta\log P = 0.23$ (the range indicated by
  the horizontal error bars). Downward-pointing arrows indicate upper
  limits.
\label{fig:giant}}       
\end{figure}

\runinhead{Overall occurrence} For giant planets the key references
are \citet{Cumming+2008} and \citet{Mayor+2011} for Doppler surveys,
and \citet{Santerne+2016} for {\it Kepler}.  These studies agree that
giant planets with periods shorter than a few years are found around
$\approx$10\% of Sun-like stars (see Figure~\ref{fig:giant}).  In
particular \citet{Cumming+2008} studied planets with a minimum mass
$m$ in the range from 0.3--10~$M_{\rm Jup}$ and $P$ from 2--2000~days.
They fitted a power-law of the form given by
Equation~\ref{eqn:powerlaw}, finding $\alpha=-0.31\pm 0.20$ and
$\beta=0.26\pm 0.10$, normalized such that 10.5\% of Sun-like stars
have such a planet.\footnote{Technically, \citet{Cumming+2008}
  calculated the fraction of stars with planets, rather than the
  average number of planets per star. The value of $C$ is $1.04\times
  10^{-3}$ when mass is measured in $M_{\rm Jup}$ and period is
  measured in days.}  They found the data to be equally well described
by a distribution uniform in $\log P$ from 2--300 days (i.e.,
$\beta=0$), followed by a sharp increase by a factor of 4--5 for
longer periods.  The {\it Kepler} data are also consistent with the
latter description \citep{Santerne+2016}. This uptick in planet
occurrence at long periods might be related to the location of the
``snow line'' in protoplanetary disks, which plays a role in the
theory of giant-planet formation via core accretion; beyond this line
there is enough snow (frozen volatiles) to pack onto a growing
protoplanet and help it to achieve the critical mass for runaway
accretion of hydrogen and helium gas \citep{Pollack+1996,Lecar+2006}.

\runinhead{Metallicity} The earliest Doppler surveys revealed the
occurrence of giant planets with periods shorter than a few years to
be a steeply rising function of the host star's metallicity
\citep{Santos+2003,FischerValenti2005}.  This too is widely
interpreted as support for core accretion theory.  The logic is that
the rapid assembly of a massive solid core --- an essential step in
the theory --- is easier to arrange in a metal-rich protoplanetary
disk.  \citet{FischerValenti2005} found their sample of
Doppler-detected giant planets to be compatible with $n\propto z^2$,
where $z$ is the iron-to-hydrogen abundance relative to the solar
value. Most recently, \citet{Petigura+2018} used {\it Kepler} data to
determine the best-fitting parameters of
\begin{equation}
\Gamma_{P,z} = \frac{\partial^2N}{\partial\log P~\partial\log z} = C\,P^\alpha z^\beta.
\end{equation}
For hot Jupiters they found $\beta = 3.4\pm 0.9$, a remarkably strong
dependence.  However, for companions more massive than 4~$M_{\rm
  Jup}$, \citet{Santos+2017} found the association with high
metallicity to be much weaker or absent, suggesting that such objects
do not form through core accretion.  \citet{Schlaufman2018} reached
the same conclusion with a sample spanning a larger range
of companion masses, and went so far as to say that companions more
massive than 10~$M_{\rm Jup}$ should not be considered planets.  The
metallicity effect is also weaker for planets smaller than Neptune
\citep{Buchhave+2012}, although for orbital periods shorter than about
10 days, even small planets are associated with elevated metallicity
\citep{Mulders+2016,Wilson+2018,Petigura+2018}.

\runinhead{Hot Jupiters} Easy to detect, but intrinsically uncommon,
hot Jupiters have an occurrence rate of 0.5--1\% for periods between 1
and 10~days. They are even rarer for periods shorter than one day
\citep{Howard+2012,SanchisOjeda+2014}.  There is a $\approx$2$\sigma$
discrepancy between the rate of 0.8--1.2\% measured in Doppler surveys
\citep{Wright+2012,Mayor+2011} and 0.6\% measured using {\it Kepler}
data \citep{Howard+2012,Petigura+2018}.  This is despite the similar
metallicity distributions of the stars that were searched
\citep{Guo+2017}.  While we should never lose too much sleep over
2$\sigma$ discrepancies, it might be caused by misclassified stars and
unresolved binaries in the {\it Kepler} sample \citep{Wang+2015}.

\runinhead{Jupiter analogs} A perennial question is whether the Solar
System is typical or unusual in some sense.  It is difficult to answer
because the current Doppler and transit surveys are only barely
sensitive to the types of planets found in the Solar System: the inner
planets are too small and the outer planets have periods that are too
long.  The most easily detected planet in an extraterrestrial Doppler
survey would probably be Jupiter; hence, a few groups have tried to
quantify the occurrence of solar-like systems by searching for
Jupiter-like exoplanets.  \cite{Wittenmyer+2016} presented the latest
effort, finding the occurrence rate to be $6.2_{-1.6}^{+2.8}$\% for
planets of mass 0.3--13~$M_{\rm Jup}$ with orbital distances from
3--7~AU and eccentricities smaller than 0.3.  Of course the rate
depends on the definition of ``Jupiter analog'', a term without a
precise meaning.  The same problem arises when trying to measure the
occurrence of ``Earth-like'' planets.

\runinhead{Long-period giants} Regarding the more general topic of
wide-orbiting giant planets, \citet{ForemanMackey+2016} measured the
occurrence of ``cold Jupiters'' with periods ranging from 2--25~years,
by searching the {\it Kepler} data for stars showing only one or two transits
over 4 years.  For planets with $R=0.4$--1.0~$R_{\rm Jup}$ they found
\begin{equation}
  \Gamma_{R,P} = \frac{\partial^2n}{\partial\log R~\partial\log P} = 0.18\pm 0.07.
\end{equation}
Integrating over the specified ranges of radius and period gives a
total occurrence rate of $0.42\pm 0.16$ planets per star.

\citet{Bryan+2016} studied the occurrence of long-period giants
conditioned on the detection of a shorter-period giant. Using
high-resolution imaging and long-term Doppler monitoring, they
searched for wide-orbiting companions to 123 giant planets with
orbital distances ranging from 0.01 to 5~AU. They found the occurrence
of outer companions to be higher than would be predicted by
extrapolating the power-law of \cite{Cumming+2008} to longer
periods. They also found $dn/d\log P$ to decline with period, unlike
the more uniform distribution observed for closer-orbiting giant
planets.  The occurrence rate was $(53\pm 5)$\% for outer companions
of mass 1--20~$M_{\rm Jup}$ and orbital distance 5--20~AU.

\runinhead{Other properties} The giant-planet population is
distinguished by other features.  Their orbits show a broad range of
eccentricities \citep[see, e.g.,][]{UdrySantos2007}.  Their occurrence
seems to fall precipitously for masses above $\approx$10~$M_{\rm
  Jup}$, at least for orbital distances shorter than a few AU.
Because of this low occurrence, the mass range from 10--80\,$M_{\rm
  Jup}$ is often called the ``brown dwarf desert''
\citep{GretherLineweaver2006,Sahlmann+2011,Triaud+2017}.  As mentioned
earlier, the inhabitants of this desert are not strongly associated
with high-metallicity stars, unlike Jovian-mass
planets\citep{Santos+2017,Schlaufman2018}.  Occasionally we find two
giant planets in a mean-motion resonance \citep{Wright+2011}.  The
rotation of the star can be grossly misaligned with the orbit of the
planet, especially if the star is more massive than about
1.2~$M_\odot$ (Triaud, this volume).  These and other topics were
reviewed recently by \citet{WinnFabrycky2015} and Santerne (this
volume).

\section{Smaller planets}
\label{sec:small}

\runinhead{Overall occurrence} About half of Sun-like stars have at
least one planet with an orbital period shorter than 100 days, and a
size in between those of Earth and Neptune.  Planet formation theories
generally did not predict this profusion of close-orbiting planets.
Indeed some of the most detailed theories predicted that close-in
``super-Earth'' or ``sub-Neptune'' planets would be especially rare
\citep{IdaLin2008}.  Their surprisingly high abundance led to new
theories in which small planets can form in short-period orbits,
rather than forming farther away from the star and then migrating
inward \citep[see, e.g.,][]{HansenMurray2012,ChiangLaughlin2013}.

Doppler surveys provided our first glimpse at this population of
planets, and then {\it Kepler} revealed it in vivid detail. For
planets with periods shorter than 50 days and minimum masses between 3
and 30\,$M_\oplus$, two independent Doppler surveys found the
occurrence rate to be $(15\pm 5)\%$ \citep{Howard+2010} and $(27\pm
5)\%$ \citep{Mayor+2011}.  For this same period range, and planets
with a radius between 2 and 4~$R_\oplus$, analysis of {\it Kepler}
data gave an occurrence rate of $(13.0\pm 0.8)\%$ \citep{Howard+2012}.
The results of these surveys are compatible, given reasonable guesses
for the relation between planetary mass and radius
\citep{Howard+2012,Figueira+2012,WolfgangLaughlin2012}.

\begin{figure}[ht!]
\includegraphics[scale=0.43,angle=0]{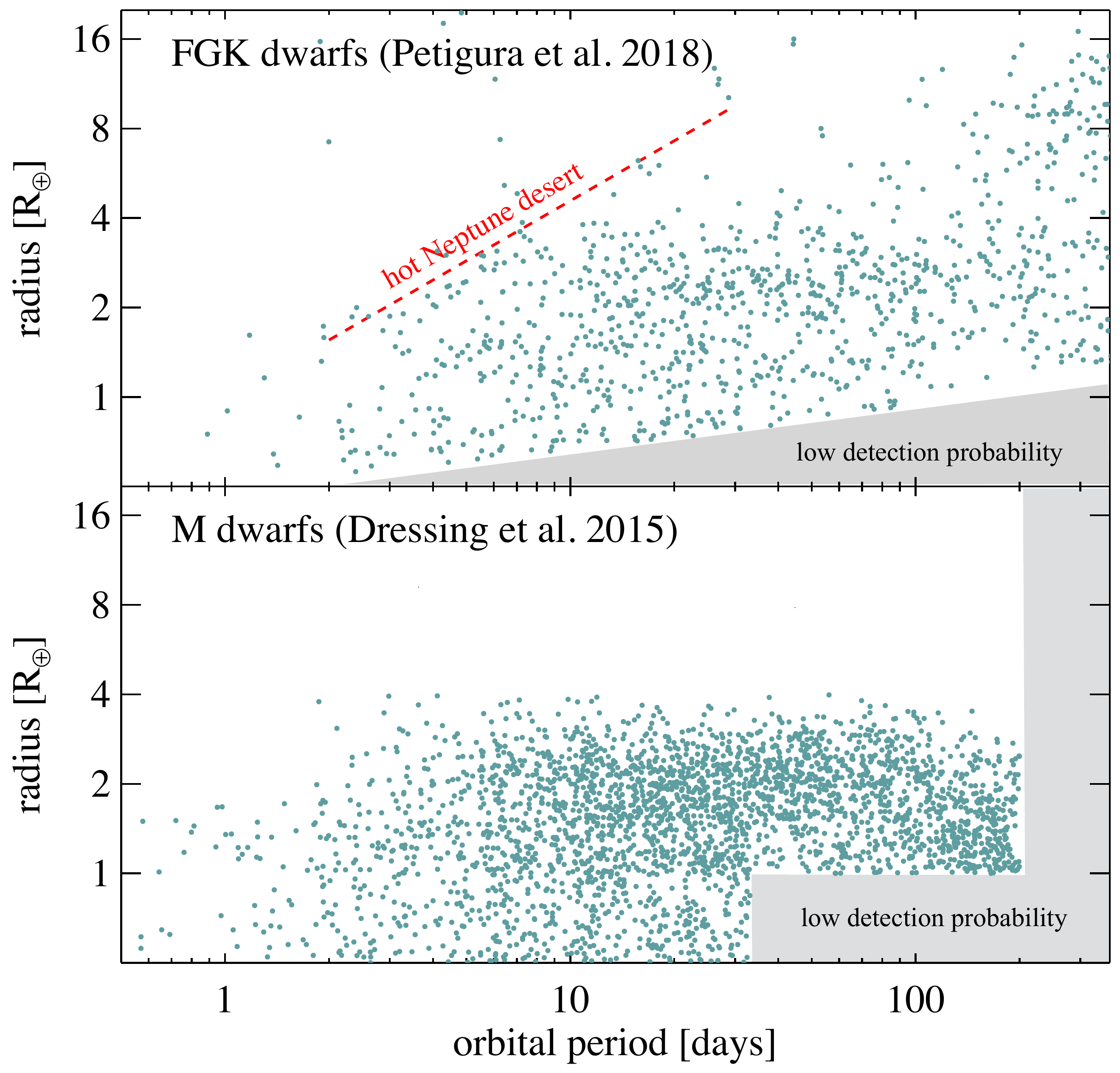}
\caption{Planet occurrence around FGK dwarfs (top) and M dwarfs (bottom)
  based on {\it Kepler} data. The blue dots represent a random sample of
  planets around $10^3$ stars, drawn from the occurrence rate densities of
  \citet{Petigura+2018} and \citet{DressingCharbonneau2015}.  Compared
  to FGK stars, the M stars have a higher occurrence of small planets
  and a lower occurrence of giant planets.
  For the M dwarfs, occurrence rates for planets
  larger than 4~$R_\oplus$ were not reported because
  only four planet candidates in that range were detected.  
\label{fig:small}}       
\end{figure}

\runinhead{Size, mass, and period} The surveys also agree that within
this range of periods and planet sizes, the occurrence rate is higher
for the smallest planets, roughly according to power laws \citep{Howard+2010,Howard+2012}:
\begin{equation}
  \frac{dn}{d\log m} \propto m^{-0.5},~~\frac{dn}{d\log R} \propto R^{-2}.
\end{equation}
For even smaller or longer-period planets, {\it Kepler} provides
almost all the available information.  Figure~\ref{fig:small} shows
some of the latest results \citep[see also][]{Fressin+2013,Burke+2015}.  The
period distribution $dn/d\log P$ rises as $\approx$$P^2$ between
1-10 days, before leveling off to a nearly constant value between
10-300 days.

\runinhead{Multiple-planet systems} Small planets occur frequently in
closely-spaced systems \citep{Mayor+2011,Lissauer+2011}, with as many
as eight planets with periods shorter than a year
\citep{ShallueVanderburg2018}.  The period ratios tend to be in the
neighborhood of 1.5--5 \citep{Fabrycky+2014}.  In units of the mutual
Hill radius,
\begin{equation}
  a_{\rm H} \equiv \left( \frac{M_{\rm in} + M_{\rm out}}{3M_\star} \right)^{1/3}
  \left( \frac{a_{\rm in} + a_{\rm out}}{2} \right),
\end{equation}
more relevant to dynamical stability, the typical spacing is
10--30 \citep{FangMargot2013}.  At the lower end of this distribution,
the systems flirt with instability \citep{Deck+2012,PuWu2015}.  A few percent of
the {\it Kepler} systems are in (or near) mean-motion resonances,
suggesting that the orbits have been sculpted by planet-disk gravitational
interactions.  These systems offer the gift of
transit-timing variations (Agol \& Fabrycky, this volume), the
observable manifestations of planet-planet gravitational interactions
that sometimes allow for measurements of planetary masses as well as orbital
eccentricities and inclinations.  Such studies and some other lines of
evidence show that the compact multiple-planet systems tend to have
orbits that are nearly circular
\citep{HaddenLithwick2014,Xie+2016,VanEylenAlbrecht2015} and coplanar
\citep{Fabrycky+2014}.

\begin{figure}
\includegraphics[scale=0.40,angle=0]{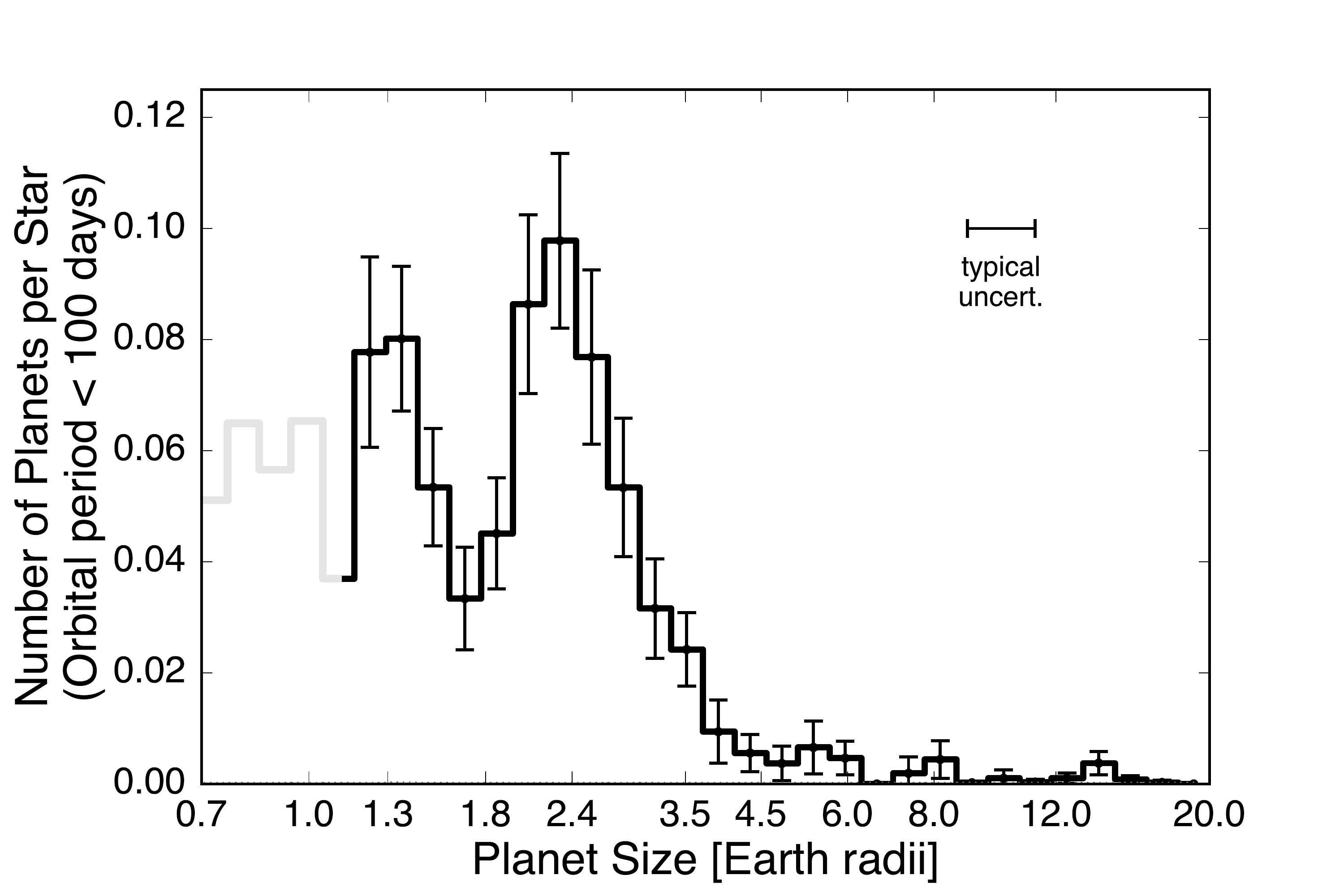}
\caption{From \citet{Fulton+2017}. Occurrence as a function of radius
  based on {\it Kepler} data, for orbital periods shorter
  than 100 days. The dip in the occurrence rate density between
  1.5--2\,$R_\oplus$ has been attributed to the erosion of
  planetary atmospheres by high-energy radiation from the star.
\label{fig:Fulton}}
\end{figure}

\runinhead{Radius gap} The radius distribution of planets with periods
shorter than 100 days shows a dip in occurrence
between 1.5 and 2\,$R_\oplus$ \citep[][see
  Figure~\ref{fig:Fulton}]{Fulton+2017,VanEylen+2017}.  Such a feature
had been anticipated based on theoretical calculations of the
photo-evaporation of the atmospheres of low-mass planets by the
intense radiation from the host star \citep{OwenWu2013,
  LopezFortney2013}.  Thus, the radius gap or ``evaporation valley'' seems to be a
precious example in exoplanetary science of a prediction fulfilled,
with many implications for the structures and atmospheres of
close-orbiting planets \citep{OwenWu2017}.

\runinhead{Hot Neptunes} As mentioned above, $dn/d\log P$ changes from
a rising function for $P\lsim10$~days to a more constant value for
$P=10$--100~days.  The critical period separating these regimes is
longer for larger planets.  The effect is to create a diagonal
boundary in the space of $\log R$ and $\log P$, above which the
occurrence is very low (see Fig.~\ref{fig:small}).  The same
phenomenon is seen in Doppler data \citep{Mazeh+2016}.  This ``hot
Neptune desert'' may be another consequence of atmospheric erosion.
Interestingly those few hot Neptunes that do exist are strongly
associated with metal-rich stars \citep{Dong+2017,Petigura+2018},
making them similar to giant planets and unlike smaller planets. The
hot Neptunes are also similar to hot Jupiters in that they tend not to
have planetary companions in closely-spaced coplanar orbits
\citep{Dong+2017}.  All this suggests the hot Neptunes and close-in
giant planets originate in similar circumstances, possibly from some
type of dynamical instability.

\runinhead{Earth-like planets} A goal with broad appeal is measuring
the occurrence rate of Earth-sized planets orbiting Sun-like stars
within the ``habitable zone'', the range of distances within which a
rocky planet could plausibly have oceans of liquid water. The {\it
  Kepler} mission provided the best-ever data for this purpose.
However, even {\it Kepler} was barely sensitive to such planets. The
number of detections was of order 10, depending on the
definitions of ``Earth-sized'', ``Sun-like'' and ``habitable
zone''. The desired quantity can be understood as an integral
\begin{equation}
  \eta_\oplus \equiv \int_{R_{\rm min}}^{R_{\rm max}} \int_{S_{\rm min}}^{S_{\rm max}}
\frac{\partial^2n}{\partial\log S~\partial\log R}~d\log S~d\log R, 
\end{equation}
where $S$ is the bolometric flux
the planet receives from the star. The integration limits are chosen to
select planets likely to have a solid surface with a temperature
and pressure allowing for liquid water. These
limits depend on assumptions about the structure and atmosphere
of the planet and the spectrum of the star \citep[see,
  e.g.,][]{Kasting+2014}.

Even if we set aside the problem of setting the integration limits,
the measurement of
\begin{equation}
\Gamma_\oplus \equiv \left.\frac{\partial^2n}{\partial\log P~\partial\log R}\right|_{P=1~{\rm yr},~R=R_\oplus}
\end{equation}
has proven difficult and may require extrapolation from measurements
of larger planets at shorter periods.  The {\it Kepler} team has
published a series of papers reporting steady advancement in the
efficiency of detection, elimination of false positives, and
understanding of instrumental artifacts.  The most recent effort to
determine $\Gamma_\oplus$ found the data to be compatible with values
ranging from 0.04 to 11.5 (see Figure~6).  Since then the {\it Kepler}
team and other groups have clarified the properties of the stars that
were searched \citep{Petigura+2017,DeCat+2015}, and the most recent
installments by \citet{Twicken+2016} and \citet{Thompson+2017}
quantified the sensitivity of the algorithms for planet detection and
validation. These developments have brought us right to the threshold
of an accurate occurrence rate for Earth-like planets.

\begin{figure}
\includegraphics[scale=.62,angle=0]{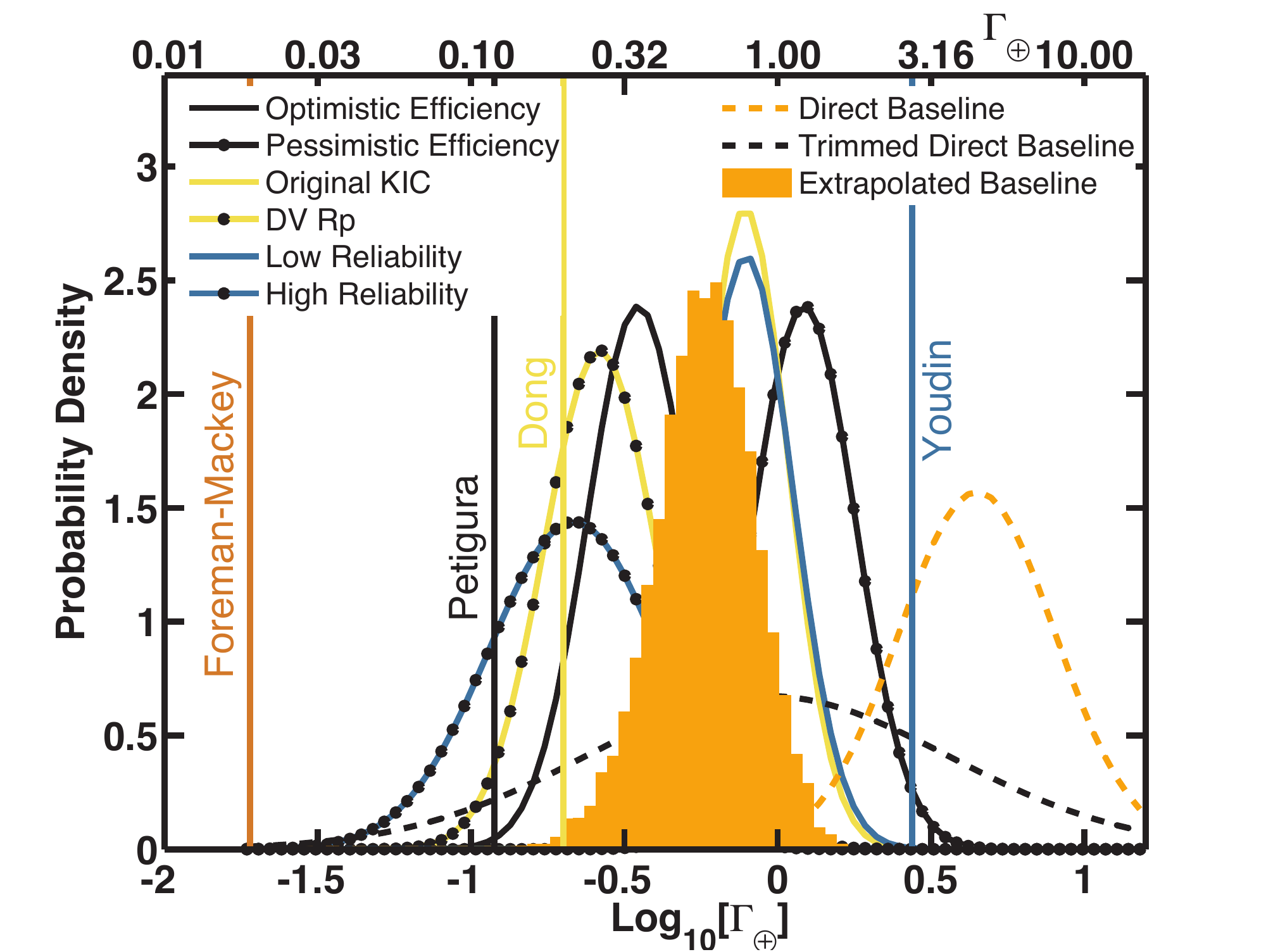}
\caption{From \citet{Burke+2015}. Estimates for $\Gamma_\oplus$ based on {\it Kepler} data.
  The orange histogram is the posterior
  probability distribution considering only the uncertainties from
  counting statistics and extrapolation.  The other curves illustrate
  the effects of some systematic errors: uncertainty in the detection
  efficiency, orbital eccentricities, stellar parameters, and
  reliability of weak planet candidates. These systematic effects led
  to a range in $\Gamma_\oplus$ spanning an order of magnitude.  The
  vertical lines show other estimates of $\Gamma_\oplus$ by
  \citet{ForemanMackey+2014,Petigura+2013,DongZhu2013} and
  \cite{Youdin2011}.}
\label{fig:BurkeFig}
\end{figure}

\section{Other types of stars}
\label{sec:other}

Almost all the preceding results pertain to main-sequence stars with
masses between 0.5 and 1.2~$M_\odot$, i.e., spectral types from K to
late F.  Stars with masses between 0.1 and 0.5~$M_\odot$, the M
dwarfs, are not as thoroughly explored, especially near the low end of
the mass range.  However these stars are very attractive for planet
surveys because their small masses and sizes lead to larger Doppler
and transit signals, and because planets in the habitable zone have
conveniently short orbital periods.

Giant planets are relatively rare around M dwarfs, at least for
periods shorter than a few years.  \citet{Cumming+2008} showed that if
planet occurrence is modeled by the functional form of
Equation~(\ref{eqn:powerlaw}), then planets with masses exceeding
0.4~$M_{\rm Jup}$ and periods shorter than 5.5 years are 3--10 times
less common around M dwarfs than around FGK dwarfs.  Similar results
were obtained by \citet{Bonfils+2013}.

On the other hand, for smaller planets over the same range of periods,
M dwarf occurrence rates exceed those of FGK dwarfs by a factor of
2--3 \citep{Howard+2012,Mulders+2015}.  This result is based on {\it
  Kepler} data, which remains our best source of information on this
topic despite the fact that only a few percent of the {\it Kepler}
target stars were M dwarfs.  Comprehensive analyses have been
performed by \citet{DressingCharbonneau2015} and \citet{Gaidos+2016}.
Their results differ in detail but agree that on average, M dwarfs
have about two planets per star with a radius in between those of
Earth and Neptune, and an orbital period shorter than 100 days (see
Figure~\ref{fig:small}).  One implication of this high occurrence rate
is that the nearest habitable-zone planets are almost surely around M
dwarfs.  Indeed, Doppler surveys have turned up two candidates for
``temperate'' Earth-mass planets around M dwarfs within just a few
parsecs: Proxima~Cen \citep[1.3~pc,][]{AngladaEscude+2016} and
Ross~128 \citep[also known as Proxima Vir; 3.4~pc,][]{Bonfils+2017}.

Some other comparisons with FGK dwarfs have been made.  Among the
similarities are that M dwarfs often have compact systems of multiple
planets \citep{Muirhead+2015,Gaidos+2016,BallardJohnson2016}, and that
high metallicity is associated with giant planet occurrence
\citep{Johnson+2010,Neves+2013}.  It remains unclear whether or not
the occurrence of smaller planets is associated with high metallicity
for M dwarfs \cite[see, e.g.,][]{GaidosMann2014}.  There is also
evidence that the planet population around M dwarfs exhibit both the
``evaporation valley'' between 1.5--2\,$R_\oplus$ and the ``hot Neptune
desert'' \citep{Hirano+2017}.

Beyond the scope of this review, but nevertheless fascinating, are the
occurrence rates that have been measured in Doppler and transit
surveys of other types of stars: evolved stars
\citep{Johnson+2010,Reffert+2015}, stars in open clusters
\citep{Mann+2017} and globular clusters
\citep{Gilliland+2000,MasudaWinn2017}, binary stars
\citep{Armstrong+2014}, brown dwarfs \citep{He+2017}, and white dwarfs
\citep{Fulton+2014,VanSluijsVanEylen2018}.

\section{Future Prospects}
\label{sec:future}

The data from recent surveys offer opportunities for progress for at
least another few years.  The ongoing struggle to measure the
occurrence rate of Earth-like planets has already been described.
Another undeveloped area is the determination of joint and conditional
probabilities; for example, given a planet with radius $R_1$ and
period $P_1$, what is the chance of finding another planet around the
same star with radius $R_2$ and period $P_2$?  Conditional rates, or
the relative occurrence of different types of systems, may be more
useful than overall occurrence rates for testing planet formation
theories.  Only a few cases have been studied, such as the mutual
radius distribution of neighboring planets \citep{Ciardi+2013,
  Weiss+2018}, and the probability for giant planets to have
wider-orbiting companions
\citep{Huang+2016,Bryan+2016,SchlaufmanWinn2016}.

New data are also forthcoming.  The stellar parallaxes soon to be
available from the ESA {\it Gaia} mission \citep{Gaia+2016} will
clarify the properties of all the {\it Kepler} stars as well as the
targets of future surveys.  Among these future surveys is the
Transiting Exoplanet Survey Satellite ({\it TESS}), scheduled for
launch in 2018 \citep{Ricker+2015}. This mission was not designed to
measure planet occurrence rates, but rather to pluck low-hanging
fruit: short-period transiting planets around bright stars. Less well
appreciated is that {\it TESS} may be superior to {\it Kepler} for
measuring the occurrence of planets larger than Neptune with periods
shorter than 10 days.  When {\it TESS} was conceived it was expected
that limitations in data storage and transmission would restrict the
search to $\approx$$10^5$ pre-selected stars, as was the case with
{\it Kepler}.  Later it became clear that entire {\it TESS} images
could be stored and transmitted with 30-minute time sampling.  As a
result, although $\approx$\,$10^5$ Sun-like stars will still be
selected for finer time sampling, it should be possible to search
millions of stars for large and short-period planets. This includes
hot and massive stars, for which comparatively little is known.
{\it TESS} should excel at finding rare, large-amplitude, short-period
photometric phenomena of all kinds.

For smaller planets around Sun-like stars, it will be difficult to
achieve an order-of-magnitude improvement over the existing data.
There is more room for advances in the study of low-mass stars, using
new Doppler spectrographs operating at far-red and infrared
wavelengths, and ground-based transit surveys focusing exclusively on
low-mass stars.  Particularly encouraging was the discovery of
TRAPPIST-1, a system of seven Earth-sized planets orbiting an
``ultra-cool dwarf'' that barely qualifies as a star
\citep{Gillon+2017}.  This system was found after searching
$\approx$50 similar objects with a detection efficiency of around 60\%
(Burdanov et al., this volume; M.\ Gillon, private communication), and
the transit probability for the innermost planet is 5\%.  This
suggests the occurrence of such systems is approximately
$(50\cdot0.6\cdot0.05)^{-1} = 0.7$.  Thus, while TRAPPIST-1 seems
extraordinary, it may represent a typical outcome of planet formation
around ultra-cool dwarfs.

In the decades to come, the domains of all the planet detection
techniques --- including direct imaging, gravitational microlensing,
and astrometry --- will begin overlapping.  Some efforts have already
been made to determine occurrence rate densities based on data from
very different techniques \citep[see,
  e.g.][]{Montet+2014,ClantonGaudi2016}.  We can look forward to a
more holistic view of the occurrence of planets around other stars,
barring any civilization-ending cataclysm.



\begin{acknowledgement}
  The author is grateful to Hans Deeg and Natalie Batalha for the
  invitation to write this review and for their editorial guidance;
  Bill Borucki, Luke Bouma, Fei Dai, Courtney Dressing, Kento Masuda, Erik Petigura,
  and Saul Rappaport, for comments on the
  manuscript; and Chris Burke, Courtney Dressing, B.J.\ Fulton, Erik
  Petigura, and Alexandre Santerne for allowing the use of the data
  and figures from their papers.
\end{acknowledgement}

\bibliographystyle{spbasicHBexo}  
\bibliography{references} 

\end{document}